\documentclass[aps,pre,twocolumn,a4paper]{revtex4-1}

\usepackage{graphicx}
\usepackage{amssymb}
\usepackage{amsmath}
\usepackage{verbatim}

\newcommand{\nn}{\nonumber}
\newcommand{\veps}{\varepsilon}

\renewcommand{\eqref}[1]{Eq. (\ref{#1})}
\newcommand{\figref}[1]{Fig. \ref{#1}}

\begin{document}

\title{The interaction between colloids in polar mixtures above $T_c$} 
\author{Sela Samin} %

\author{Yoav Tsori} 
\affiliation{Department of Chemical Engineering and The Ilse Katz Institute for
Nanoscale
Science and Technology, Ben-Gurion University of the Negev, 84105 Beer-Sheva,
Israel.}

\date{\today}

\begin{abstract}
We calculate the interaction potential between two colloids immersed in an aqueous
mixture containing salt near or above the critical temperature. We find an
attractive interaction far from the coexistence curve due to the combination of
preferential solvent adsorption at the colloids' surface and preferential ion
solvation. We show that the ion-specific interaction  strongly depends on the
amount of salt added as well as on the mixture composition. Our results are in
accord with recent experiments. For a highly antagonistic salt of
hydrophilic anions and hydrophobic cations, a repulsive interaction at an
intermediate inter-colloid distance is predicted even though both the
electrostatic and adsorption forces alone are attractive.
\end{abstract}


\maketitle

\section{Introduction}

Charged surfaces in liquid media are ubiquitous in soft matter and the interaction
between such surfaces has been studied extensively. In the seminal theory of
Derjaguin, Landau, Verwey,
and Overbeek (DLVO) the total interaction potential is the additive combination
of the attractive van der Waals interaction and an electrostatic repulsive
screened Coulomb potential \cite{dlvo1,dlvo2}. In the DLVO theory the liquid
between the surfaces is a homogeneous dielectric medium and only enters the
electrostatic potential through its permittivity $\veps$.
The situation is different in solvent \emph{mixtures}, where the structure of the solvent
may vary in space due to gradients in electric field and ion density.
Several authors explored experimentally
the interaction between charged surfaces in mixtures
\cite{Beysens1985,Beysens1991,Bonn2009,Bechinger2008,nellen2011}. In a
homogeneous binary mixture of water and 2,6-lutidine close to the demixing
temperature, a reversible flocculation of suspended colloids has been observed
\cite{Beysens1985}. The flocculated region in the phase diagram is suppressed by the
addition of salt \cite{Beysens1991}. These experiments were interpreted in terms of the
preferential adsorption of one of the solvents
on the colloid surface in accord with wetting theory
\cite{Beysens1998,Evans2009}. More recently, it was suggested that the selective
solvation of ions in the two solvents is important in these experiments
\cite{efips_epl,onuki2011} as well as charge-regulation effects
\cite{onuki2011}.

Recently, direct measurement of the interaction potential between a colloid and
a charged wall was performed in water--2,6-lutidine mixtures below the lower
critical solution temperature (LCST) \cite{Bechinger2008,nellen2011}. In these
experiments the addition of salt greatly influenced the range and strength of
the interaction \cite{nellen2011}, leading to an attractive interaction far from
the critical temperature $T_c$. Addition of salt has a remarkable effect on the
flocculation of colloids in a critical mixture of water--3-methylpyridine, where
it reduces dramatically the onset temperature of the flocculation
\cite{Bonn2009}. Preliminary experiments \cite{bechinger2011} suggest that the
variation of the mixture composition also has a pronounced effect on the
interaction.

A theoretical effort has been made to
incorporate ionic solvation effects in polar mixtures
\cite{onuki2011,efips_epl,bier2011,ciach2011,andelman_jpcb2009,*benyaakov2011}.
The solvation energy of an ion arising from the ion-dipole interaction and
sometimes other specific interactions with the polar solvent strongly depends on
the solvent and the chemical nature of the ion \cite{israelachvili,
marcus_cation, marcus_anion}. The total solvation energy of an ion when it is
transfered from one solvent into another (the Gibbs transfer energy) is in many
cases much larger than the thermal energy, especially if one solvent is water
and the other is an organic solvent \cite{ marcus_cation, marcus_anion,
persson1990}. In addition, for a given solvent, the solvation energy of cations
and anions is generally different due to their different size and specific
chemical interactions with the solvent, such as Lewis acid-base interaction.
Thus, the magnitude and sometimes even the sign of the Gibbs transfer energy for
cations and anions may differ in various salts and solvents
\cite{persson1990,marcus_book1}.

In bulk binary mixtures, the addition of salt modifies the coexistence curve
\cite{seah1993,Fuess1999,onuki:021506,PhysRevE.82.051501,onuki_review}. When a
salt-containing mixture is confined between charged surfaces, the influence of
ionic solvation on the surface interaction in miscible solvents has been studied
in Ref. \onlinecite{andelman_jpcb2009,*benyaakov2011}, while for partially
miscible mixtures close to the coexistence curve it was explored in Ref.
\onlinecite{efips_epl}. In the latter case it was shown that preferential
solvation alone can qualitatively modify the interaction between similarly
charged surfaces, resulting in a strong long-range attraction.

In order to account for the recent experimental findings in critical mixtures
\cite{nellen2011},
Bier \textit{et. al.} \cite{bier2011} had employed a linear theory that includes
both ion solvation and solvent adsorption on the confining surfaces. Recently,
Okamoto and Onuki \cite{onuki2011} also included in the theory solvent dependent
charge-regulation, and studied the interaction induced by prewetting and wetting
transitions close to the coexistence curve.

In this paper we theoretically investigate how the interplay between ion
solvation and solvent adsorption affects the interaction between surfaces in the
nonlinear regime. We show that a strong attraction exists even far from $T_c$
due to the nonlinear coupling between solvation and adsorption. These effects
are strongly ion-specific and hence depend on the salt properties and
concentration. We highlight the qualitative features of the theory 
and give several new predictions. 

We focus on the regime far from the prewetting and wetting transitions and
neglect charge-regulation, assuming large surface ionization. Thus, we do not
expect any discontinuities in physical quantities such as pressure and
adsorption. We concentrate on electrostatics and ion-solvation forces and hence
ignore the van der Waals interaction (which is relatively small in the current
range of parameters).

The organization of this paper is as follows. In Sec. \ref{sec_model}
we present a coarse-grained model of a salt-containing binary mixture, taking
into account the specific ion-solvent and surface-solvent interactions. In Sec.
\ref{sec_results} we present results for critical and off--critical mixture
compositions and show how ion specific effects can be used to tune the
inter--colloidal potential. Conclusions are given in Sec. \ref{sec_conclusions}.

\section{Model}
\label{sec_model}

The surfaces of the two colloids are modeled as parallel plates of area $S$ separated by a
distance $D$ and carrying a uniform charge density $e\sigma_{L}$ and
$e\sigma_{R}$ per unit area, where $e$ is the elementary charge and the indices
$L$ and $R$ denote the left and
right plates, respectively. An aqueous binary mixture containing salt is
confined between the plates. $\phi$ is the volume fraction of the water
($0\leq \phi\leq 1$), while the densities of the positive and negative ion are
denoted by
$n^+$ and $n^-$, respectively. Given their low density and small size, the
volume fractions of the ions are neglected. 

For the fluid free energy we have $F_b=\int f_b {\rm d}\mathbf{r} $, where $f_b$
is the bulk free energy density: \cite{onuki:021506,tsori_pnas_2007} 
\begin{align}
\frac{f_b}{T}&=f_{m}(\phi)+\frac{C}{2}|\nabla \phi|^2 +\frac{1}{2T}
\veps(\phi)(\nabla
\psi)^2 \nn \\ &+n^+\left(\log (v_0n^+)-1\right)+n^-\left(\log (v_0n^-)-1\right)
\nn\\
&-(\Delta u^+n^++\Delta u^-n^-)\phi.
\label{eq:fmions}
\end{align}
Here the Boltzmann constant is set to unity and $T$ is the thermal energy.
$v_0=a^3$ is the molecular volume of the solvent molecules where $a$ is the
linear dimension of the molecules. The first term in \eqref{eq:fmions} is the
mixture free energy \cite{safran}:
\begin{equation}
v_0f_m=\phi\log(\phi)+
(1-\phi)\log(1-\phi)+\chi\phi(1-\phi),
\end{equation}
where $\chi \sim 1/T $ is the Flory interaction parameter. In this symmetric
model the critical composition is given by $\phi_c=1/2$. The square-gradient term in
\eqref{eq:fmions} takes into account the free energy increase due to composition
inhomogeneities \cite{safran}, where $C$ is a positive constant. The third term
in \eqref{eq:fmions} is the electrostatic energy density, where $\psi$ is the
electrostatic potential. This energy depends on the composition through the
constitutive relation for the dielectric constant: $\veps=\veps(\phi)$. For
simplicity we use the linear form: $\veps(\phi)=\veps_c+(\veps_w-\veps_c)\phi$,
where $\veps_w$ and $\veps_c$ are the water and cosolvent permittivities,
respectively. The second line of \eqref{eq:fmions} is the ideal-gas entropy of
the ions, valid at low densities, while the third line is the ion solvation in
the mixture. In the simple bilinear form we employ, the strength of selective
solvation is proportional to the coefficients $\Delta u^+$ and $\Delta u^-$ for
positive and negative ions, respectively \cite{tsori_pnas_2007}. In this form the Gibbs
transfer
energies of the ions are $T\Delta u^\pm(\phi_2-\phi_1)$, where $\phi_1$ and
$\phi_2$ are the compositions of two solvents.

In addition to the bulk term, one adds the contribution of the confining
surfaces to the free energy: $F_s=\int f_s {\rm d}S$, where $f_s$ is the surface
free energy density given by:
\begin{equation}
f_s=e\sigma\psi_s+\Delta\gamma\phi_s,
\label{eq:fsions}
\end{equation}
where $\psi_s$ and $\phi_s$ are the electrostatic potential and mixture
composition at the surface, respectively. \eqref{eq:fsions} includes a short
range interaction between the surface and liquid which is linear in $\phi$,
neglecting higher order terms. The surface field $\Delta\gamma$ measures the
difference between the surface-water and surface-cosolvent surface tensions.

The fluid is in contact with a matter reservoir having composition $\phi_0$ and
ion densities $n_0^\pm$. Hence, the appropriate thermodynamic potential to
extremize is the grand potential $\Omega$:
\begin{equation}
\Omega=F_b+F_s-\int[\mu \phi+\lambda^+n^++\lambda^-n^-]{\rm d} \mathbf{r},
\end{equation}
where $\mu$ and $\lambda^\pm$ are the Lagrange multipliers for the composition
and ions, respectively. They can be identified as the chemical potentials of the
species in the reservoir: $\mu=\mu_0(\phi_0,n^\pm_0)$ and
$\lambda^\pm=\lambda^\pm_0(\phi_0,n^\pm_0)$.

The governing equations are obtained by minimization of $\Omega$, leading to
the Euler-Lagrange equations
\begin{align}
\label{eq:npm}
\frac{\delta \Omega}{\delta n^\pm}=&\frac{\pm e \psi}{T}+\log (v_0n^\pm)-\Delta
u^\pm\phi-\lambda^\pm=0, \\
\label{eq:pb}
\frac{\delta \Omega}{\delta \psi}=&\nabla \cdot (\veps(\phi)\nabla
\psi)+(n^+-n^-)e=0,\\
\label{eq:comp}
\frac{\delta \Omega}{\delta \phi}=&-C
\nabla^2\phi+\frac{\partial f_m}{\partial \phi}-\frac{1}{2T}
\frac{d\veps}{d\phi}(\nabla\psi)^2 \nn \\
&-\Delta
u^+n^+-\Delta u^-n^--\mu=0.
\end{align}
The first equation yields the Boltzmann distribution for the ion density
\begin{align}
\label{eq:boltz}
n^\pm&=v_0^{-1}{\rm e}^{\lambda^\pm}{\rm e}^{\mp e \psi/T+\Delta u^\pm\phi}.
\end{align}
Note the dependence on the mixture composition in this equation. The ion
distributions are inserted into \eqref{eq:pb} to obtain a modified
Poisson-Boltzmann equation with the boundary conditions $-\mathbf{n} \cdot
\nabla
\psi_{L,R} = e\sigma_{L,R}/\veps$. ${\bf n}$ is the outward unit vector
perpendicular to the surface. The governing equation for the composition
[\eqref{eq:comp}] is supplemented by the boundary conditions $\mathbf{n} \cdot
\nabla \phi_{L,R} =
\Delta\gamma_{L,R}$. Assuming that $\psi=0$ in the reservoir, the Lagrange
multipliers for the different species are:
\begin{align}
\label{eq:mu0}
\mu_0&=\frac{\partial f_m}{\partial \phi}(\phi_0)
-\Delta u^+n_0^+-\Delta u^-n_0^- \\
\label{eq:lam0}
\lambda_0^\pm&=\log(v_0 n_0^\pm)-\Delta u^\pm\phi_0
\end{align}

In our effectively one dimensional system, solution of the Euler-Lagrange
equations yields the
density profile $\phi(z)$ and potential $\psi(z)$ from which we calculate
all relevant quantities. Of particular interest is the pressure $P_n$ exerted on the
surfaces by the liquid. $-P_n$ is the normal component of the Maxwell
stress tensor and it is given by \cite{landau2,andelman_jpcb2009}:
\begin{align} 
\label{eq:pi} &\frac{P_{n}}{T}=\frac{C}{2}|\nabla \phi|^2
-C\phi\nabla^2 \phi+\phi\frac{\partial f_{m}}{\partial \phi}-f_{m}+n^+(1-\Delta
u^+\phi) \nn \\ &+n^-(1-\Delta u^-\phi)-\frac{1}{2T}\left( \phi
\frac{d\veps}{d\phi}+\veps\right)(\nabla \psi)^2
\end{align}
In a planar geometry $P_n$ is the $zz$ component of the stress tensor and is
uniform in mechanical equilibrium. The net pressure on the plates is given by
the osmotic pressure $\Pi=P_{zz}-P_b$, where $P_b$ is the bulk pressure. The
interaction potential between the plates $U(D)$ is calculated from the osmotic
pressure through
\begin{align}
U(D)=-S\int_{\infty}^{D} \Pi (D') {\rm d} D'.
\end{align}

The excess surface adsorption in a planar geometry is defined as
usual by
\begin{align}
\Gamma=\frac{1}{D}\int_{0}^{D}[\phi(z)-\phi_0] {\rm d} z.
\end{align}
This quantity is measurable experimentally and accounts for the total excess of
fluid between the plates relative to the bulk.

\section{Results}
\label{sec_results}

In the following we examine the interaction potential $U(D)$ for a binary
mixture containing $10$mM of salt at temperatures larger than the critical
temperature, $T>T_c$.

\subsection{Colloidal interaction in a mixture at a critical composition}

We start with a reservoir at a critical composition, $\phi_0=\phi_c$, confined by two
hydrophilic ($\Delta\gamma_{R,L}>0$) colloids. We first
consider the symmetric case where both colloids are identical:
$\Delta\gamma_{L}=\Delta\gamma_{R}$ and $\sigma_{L}=\sigma_{R}$. We set
$\sigma_{L,R}=-\sigma_{sat}$, where $\sigma_{\rm sat}=\sqrt{8 n_0/(\pi l_B)}$
and $l_B=e^2/(4\pi\veps(\phi_0)T)$ is the Bjerrum length. $\sigma_{\rm sat}$ is
used because the effective surface charge saturates to this value in the high
bare charge limit \cite{Trizac2002}. Furthermore, we assume symmetric
preferential solvation, \textit{i.e.} $\Delta u^+=\Delta u^-$, and in this case
the solvation asymmetry parameter, defined as $\Delta u^{d} \equiv \Delta
u^+-\Delta u^-$, vanishes: $\Delta u^d=0$. 

The parameters we use are such that in the resulting profiles
$\psi(z;D)$ and $\phi(z;D)$ the conditions $ e \psi (z;D)/T\gtrsim1$ and $\Delta
u^\pm (\phi (z;D)-\phi_0)\gtrsim1$ hold. Hence, a linearization of \eqref{eq:npm} 
is not expected to be accurate \cite{bier2011,onuki2011}.
In fact, we show below that it is the nonlinear coupling between the
governing equations that is responsible to an attractive interaction far from
$T_c$.
\begin{figure}[!t] 
\includegraphics[width=8.5cm,clip]{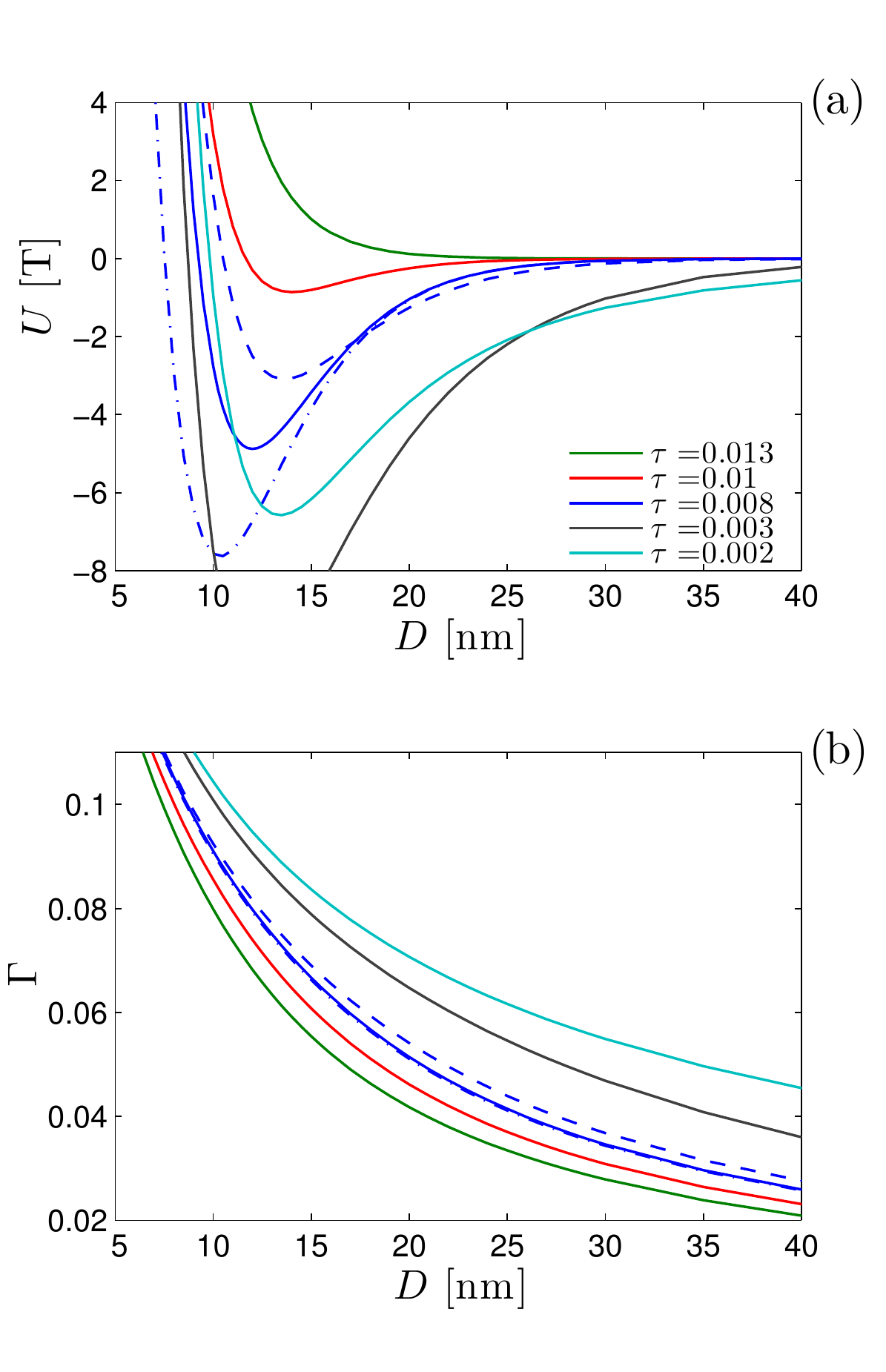}
\caption{(a) The interaction potential $U(D)$ between two colloids at a distance
$D$ at different temperatures $\tau\equiv T/T_c-1>0$ immersed in a mixture at a
critical composition ($\phi_0=\phi_c$). $U(D)$ becomes attractive as $\tau$
decreases.
Here $n_0=10$mM and $\Delta\gamma_{R,L}=0.1T/a^2$, corresponding to
about $3.4$ mN/m. For the solid curves, $\Delta u^\pm=4$ and the surfaces have
the same charge $\sigma_{L,R}=-\sigma_{sat}$. Dash-dot curve: the same as the
solid curve for $\tau=0.008$ except that
$\sigma_{L}=3\sigma_{R}=-1.5\sigma_{sat}$. Dashed curve: the same as for
$\tau=0.008$ except that $\Delta u^-=8$. (b) The corresponding excess surface
adsorption $\Gamma$. In this and in other figures, as an approximation of a
water--2,6-lutidine mixture we used $T_c=307.2$K, $v_0=3.9\times
10^{-29}$m$^3$, $C=\chi/a$ \cite{safran}, $\veps_{\rm 2,6-lutidine}=6.9$ and
$\veps_{\rm water}=79.5$, and
the surface area is taken to be $S=0.01\mu$m$^2$.} 
\label{fig1} 
\end{figure}

The solid curves of \figref{fig1}(a) show $U(D)$ as a function of the reduced
temperature $\tau\equiv T/T_c-1$ for $T>T_c$. Far above $T_c$ $U(D)$ is purely repulsive;
as the temperature is decreased toward $T_c$ the interaction becomes attractive. The
change in the shape of $U(D)$ originates from the interplay between surface field, the
solvation induced attraction, and the electrostatic repulsion between the
plates. The attraction between the plates is stronger down to $\tau=0.003$
and then becomes less attractive for $\tau=0.002$.
The attraction range of a few tenths of nm, its strength of a
few $T$, as well as the temperature dependence are
similar to those observed in recent experiments \cite{nellen2011,bechinger2011}.

The dashed and dash-dot curves in \figref{fig1}(a) are the same as the solid
curve for $\tau=0.008$ except for one parameter. The effect of charge asymmetry
is shown by the dash-dot curve, where the total charge was kept constant, but
$\sigma_{L}=3\sigma_{R}$. Here the attraction is stronger, as in the classic 
Poisson-Boltzmann theory \cite{BenYaakov2010}, since the electrostatic repulsion
between surfaces is weaker for the asymmetric case. The dashed curve shows that
increasing $\Delta u^-$ by 4 ($\Delta u^{d}=-4$) results in a slightly weaker
interaction. In the linear theory \cite{bier2011,onuki2011}, solvation coupling only enters as a
term $\propto \Delta u^{d}$. Thus, by comparison to the solid curve in \figref{fig1}(a)
where $\Delta u^{d}=0$, it is clear that in the nonlinear regime this type of
contribution is smaller than contributions proportional to $\Delta u^\pm$.

The Gibbs adsorption $\Gamma$ corresponding to the curves in \figref{fig1}(a) is
shown in \figref{fig1}(b). When the surface separation decreases $\Gamma$
increases as the adsorbed fluid layers near the walls merge. Furthermore, since
the density of the ions also increases, more fluid is drawn to the walls
because of the solvation interaction. $\Gamma$ increases as the adsorbed fluid
layer thickness, which is comparable to the bulk correlation length, increases
closer to $T_c$. 

The parameters we use are close to those used in 
experiments in a salt containing water--2,6-lutidine mixture below the LCST
\cite{nellen2011,bechinger2011}. In these experiments it is reasonable to assume
that the surfaces have different charge densities because of their
different chemical nature. Therefore, below we use
$\sigma_{L}=3\sigma_{R}=-1.5\sigma_{sat}$ for hydrophilic surfaces. The value of 
$\Delta \gamma$ was chosen arbitrarily since there is no molecular theory to predict it
accurately. Furthermore, in water--2,6-lutidine mixtures the anions are
expected to favor the water environment more than the cations, since
2,6-lutidine is a Lewis base. This is supported by data of Gibbs transfer
energies of ions in water-pyridine mixtures \cite{persson1990}. 2,6-lutidine is
a structural analog of pyridine and hence on the basis of Ref.
\onlinecite{persson1990} we took $\Delta u^+=4$ and $\Delta u^-=8$. We stress
that over a wide range of values for $\sigma_{L,R}$ and positive $\Delta
\gamma_{L,R}$, the results do not change qualitatively. This is also true for
the solvation parameters as long as the ions are hydrophilic, $\Delta
u^\pm>0$. Note that the molecular volumes of the
two components differ substantially in the real mixture leading to
$\phi_c\neq1/2$. Furthermore, our simplified model mixture has an upper critical
solution temperature, whereas experiments are performed below the LCST in
water--2,6-lutidine mixtures. Thus, one should interpret results
in terms of the absolute distance from $T_c$.

\begin{figure}[!tb] 
\includegraphics[width=8.5cm,clip]{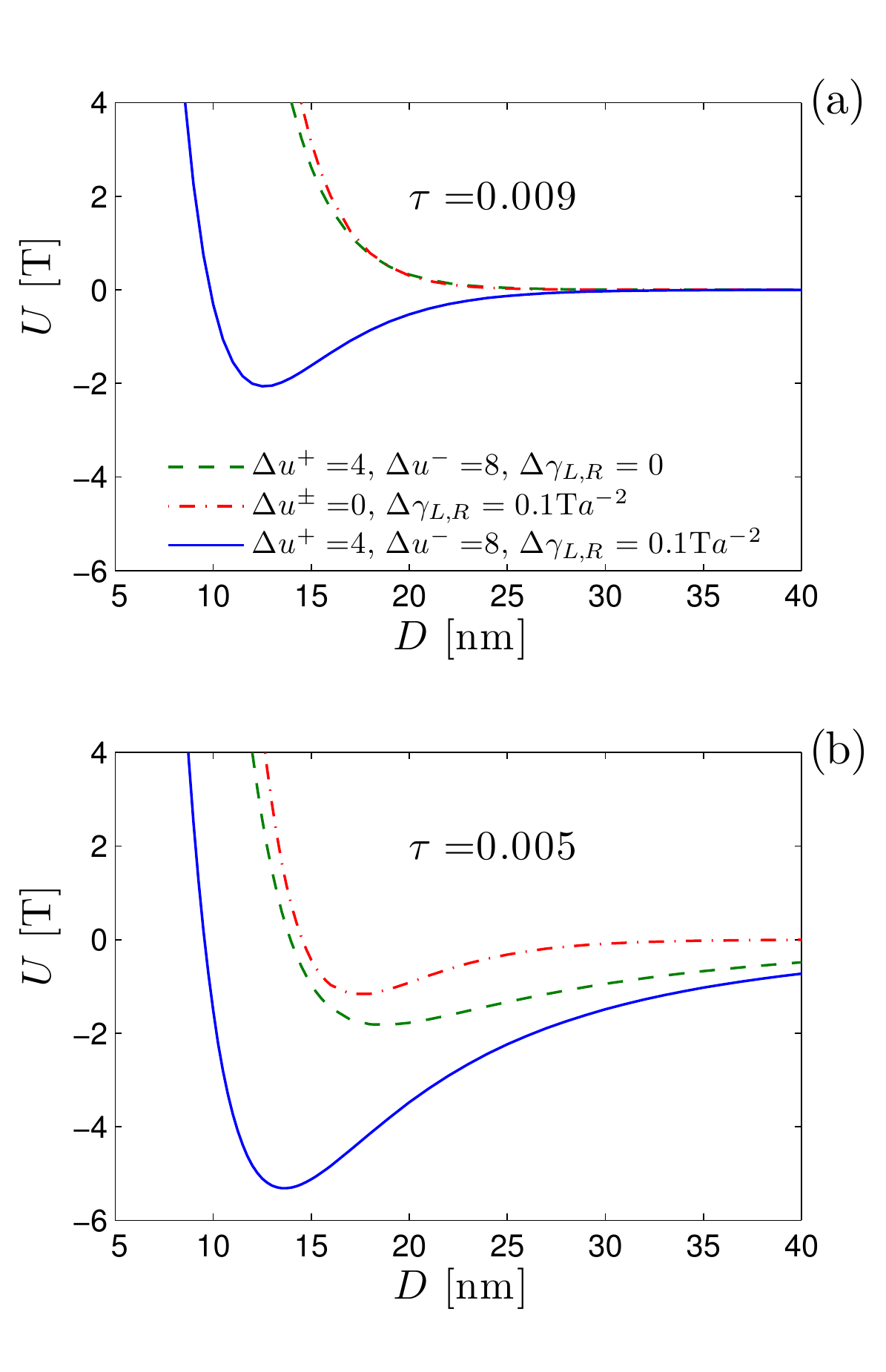}
\caption{The effect of removing preferential solvation or short-range chemical
interactions on the potential $U(D)$ between the colloids at temperatures given
by (a) $\tau=0.009$ and (b) $\tau=0.004$. The dashed and dash-dot curves show $U(D)$ when
either the surface chemical affinity or preferential solvation parameters are zero,
respectively. In the solid curves both short-range chemical
preference and solvation are included. These two interactions are clearly non additive as
the solid curve is not the sum of the dashed and dash-dotted lines.}
\label{fig2} 
\end{figure}

$U(D)$ can be attractive when only surface fields are present ($\Delta u^\pm=0$,
$\Delta \gamma_{L,R}\neq0$) due to critical adsorption
\cite{Fisher1978,Fisher1981,Krech1997,Gambassi2009,Kiselev2004}. When only
preferential solvation exists ($\Delta u^\pm\neq0$, $\Delta
\gamma_{L,R}=0$), we have shown recently \cite{efips_epl} that fluid enrichment
near the surface, induced by solvation, leads to an attractive interaction
near $T_c$. Interestingly, as is shown in \figref{fig2}, the attraction between the
surfaces is significantly altered by the coupling of solvation and surface fields.

In part (a) we highlight this idea by choosing parameters such that far from
$T_c$, removing either the surface tension (dashed curve) or preferential
solvation (dash-dot curve) the interaction is purely repulsive. When both forces
are present the result is surprisingly a strong attractive interaction between the
colloids (solid
curve). \figref{fig2}(b) shows that closer to $T_c$, an attractive interaction
of similar magnitude exists when either force is missing, but the interaction is
unexpectedly much stronger when both are present. Thus, in the nonlinear regime,
solvation and adsorption forces are non additive and their coupling can account
for attractive interactions far from $T_c$.
\begin{figure}[t!]
\includegraphics[width=8cm,clip]{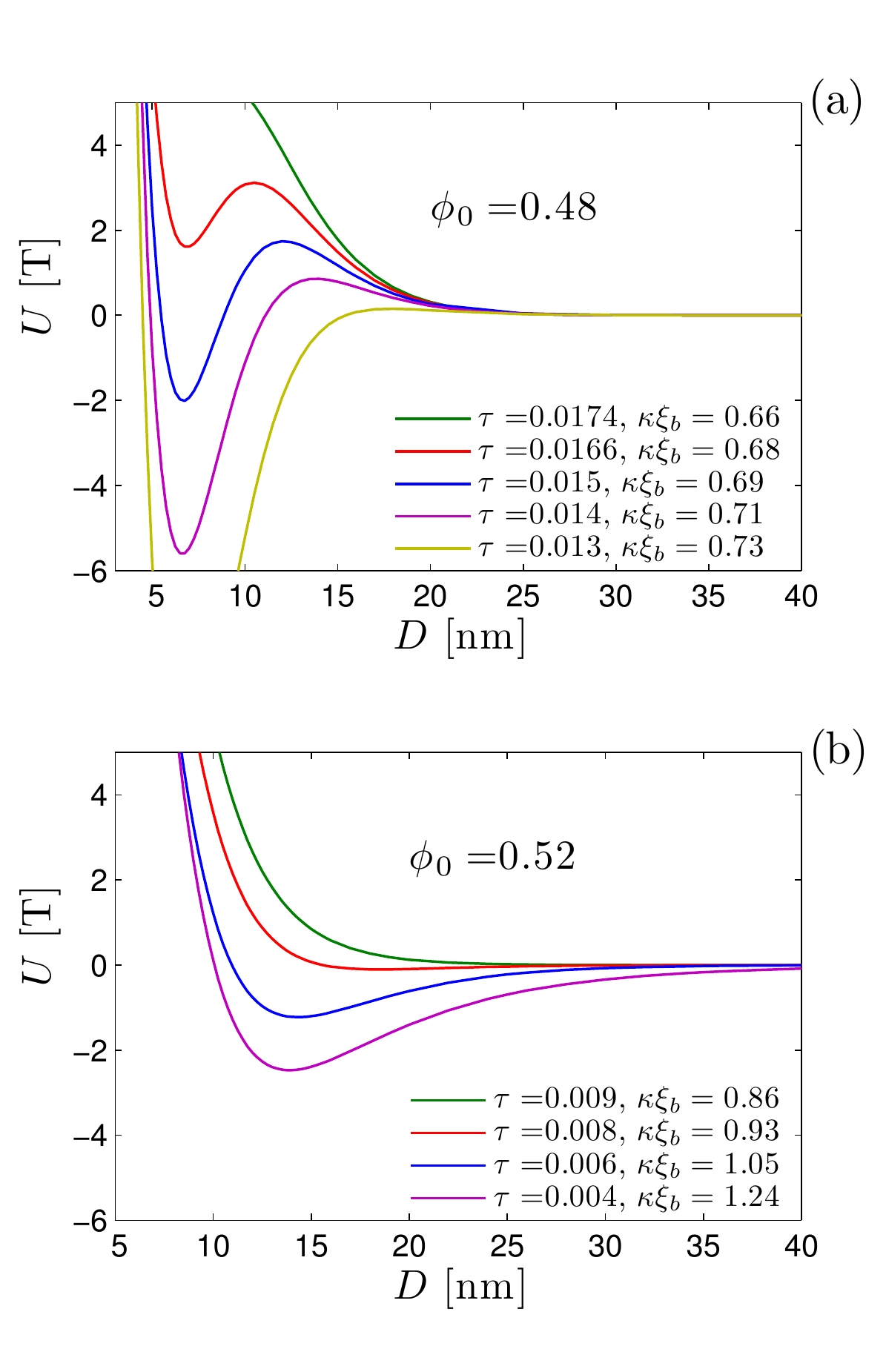}
\caption{Interaction potentials at different temperatures $\tau$, (a) for
$\phi_0=0.48<\phi_c$ with $\Delta\gamma_{R,L}=0.1T/a^2$ and (b) for
$\phi_0=0.52>\phi_c$ with $\Delta\gamma_{R,L}=0.4T/a^2$. The onset temperature
for attraction is higher for $\phi_0<\phi_c$ due to preferential solvation. In
(a), at intermediate temperatures the potential has metastable states. Here we used 
$\Delta u^+=4$, $\Delta u^-=8$ and $\sigma_{L}=3\sigma_{R}=-1.5\sigma_{sat}$. 
We took the average ion density to be $n_0=10$mM leading to $\kappa$ values of
$\kappa\simeq 2.69$nm.} 
\label{fig3} 
\end{figure}

\subsection{Surface interaction in an off-critical mixture}

The interaction potential between two surfaces for off-critical mixture
compositions at different temperatures is shown in \figref{fig3}. The legend
also shows the dimensionless parameter $\kappa\xi_b$ corresponding to $\tau$,
where $\kappa=\sqrt{8\pi l_Bn_0}$ is the Debye wave number and $\xi_b$ is the
bulk correlation length in the absence of ions, defined as
$\xi_b(\tau)=\sqrt{\chi
a^2/\left(\partial^2f_m(\phi_c)/\partial\phi^2\right)}$. The combination
$\kappa\xi_b$ is useful for comparison with experimental results.
Experimentally, it was found that ions do not modify the correlation length
significantly \cite{nellen2011}.

\begin{figure}[!t] 
\includegraphics[width=8cm,clip]{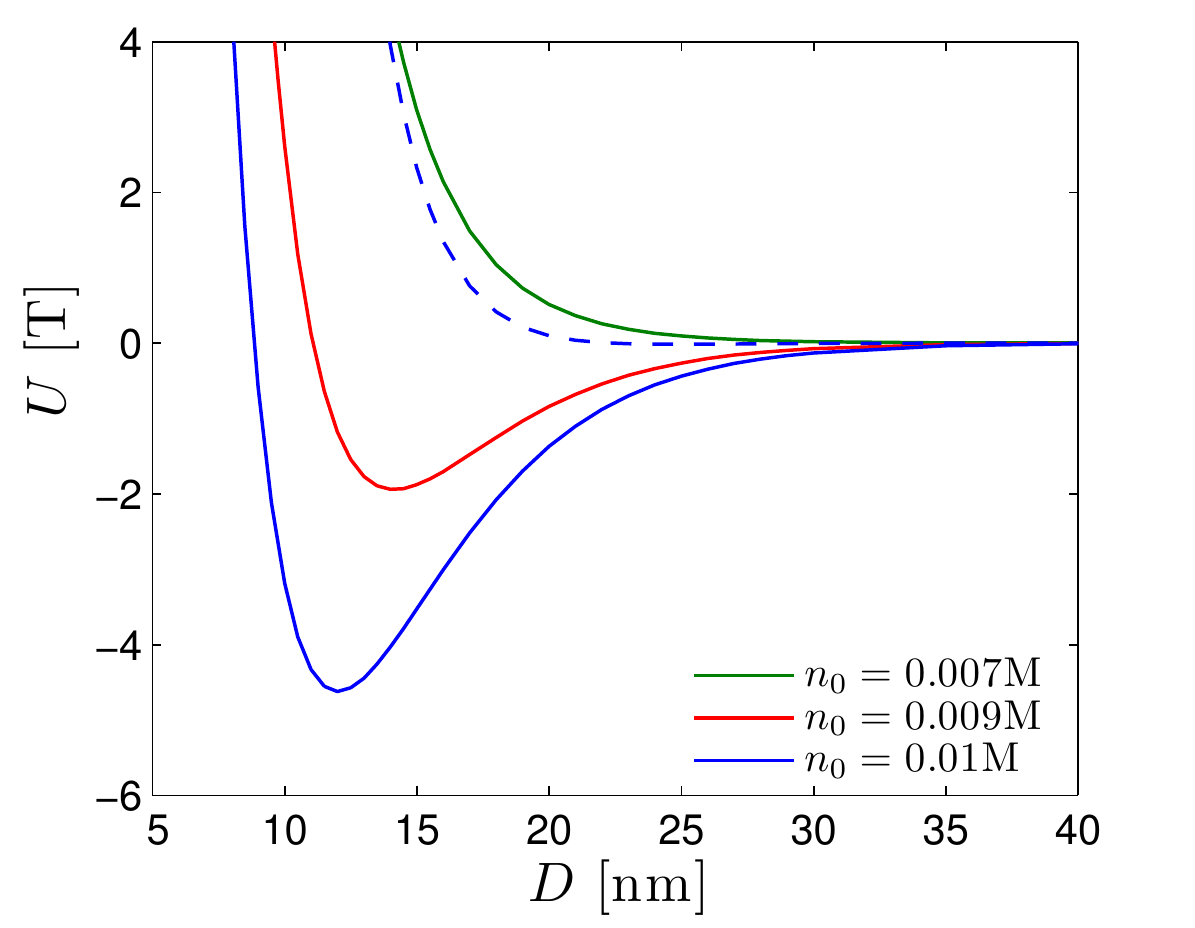}
\caption{Solid curves show the dependence of $U(D)$ on the mixture salt
concentration $n_0$ with $\Delta u^+=4$ and $\Delta u^-=8$; colloidal attraction
increases with the addition of salt. The attraction is weak for a salt
concentration of $n_0=0.01$M but with $\Delta u^\pm=0$ (dashed curve). Here
$\phi_0=\phi_c$, $\tau=0.008$,
$\Delta \gamma_{L,R}=0.1$ and $\sigma_{L}=3\sigma_{R}=-1.5\sigma_{sat}$.}
\label{fig4}
\end{figure}

As is seen in \figref{fig3} for $\phi_0\neq\phi_c$, the curve $U(D)$ becomes
attractive closer to $T_c$, similar to the $\phi_0=\phi_c$ case, but the
temperature range at which this occurs is different. The attractive interaction
is stronger for
$\phi_0<\phi_c$ [\figref{fig3} (a)] and thus the temperature range above $T_c$
at which
it occurs is larger; compare with $\tau$ values in
\figref{fig2}. Note that in \figref{fig3} (a) several curves have metastable
states and that the interaction is repulsive at long range. The attractive
interaction is weaker for $\phi_0>\phi_c$ [\figref{fig3} (b)]. Here, we had to
use a larger value of water-surface interaction in order to obtain a potential depth
similar to the one which can be resolved in experiments (few T's)
\cite{nellen2011}. 

In order to understand the dependence on reservoir composition we insert
\eqref{eq:lam0} for $\lambda_0^\pm$ into the Boltzmann distribution
\eqref{eq:boltz} to obtain 
\begin{align}
\label{eq:boltz0}
n^\pm&=n_0{\rm e}^{\mp e \psi/T+\Delta u^\pm(\phi-\phi_0)}.
\end{align}
This equation suggests that the ion density near the plates increases when
$\phi_0$ is reduced, which is verified numerically. This leads to a decrease of
the osmotic pressure in the system since when $\Delta u^\pm\phi>1$ the solvation
contribution to $\Pi$ overcomes the ions entropic repulsion [cf.
\eqref{eq:pi}]. Thus, the coupling between the mixture composition and the ion
density is important for the interaction between the surfaces. Due to the
ion-solvent coupling the attraction is increased when salt is added. This effect
is shown by the solid curves in \figref{fig4} and is seen in experiment
\cite{Bonn2009,bechinger2011}. The dashed curve in \figref{fig4} shows $U(D)$
for the highest salt concentration but in the absence of ion-solvent coupling
($\Delta u^\pm=0$). In this case the attraction is very weak, indicating
that the decrease in Debye length when salt is added is not significant for the
potential at the given parameters.

In general, attraction occurs up to a temperature window $\Delta T\equiv T-T_c
\approx 5$K above $T_c$, quite far from
$T_c$ but not as large as $\Delta T \approx 10$K recently observed in
experiments. This is reasonable given
the simplified mixture model, difference in geometry and many approximations of
unknown quantities, \textit{e.g.} $\sigma$, $\Delta\gamma$, and $S$. Also, in
experiments the asymmetric binodal
curve of water--2,6-lutidine mixtures is much ''flatter'' near $T_c$
compared to the binodal in the symmetric Flory-Huggins model we use.

Better agreement with experiments is achieved when the values of $\kappa\xi_b$
are compared. Attraction is observed experimentally when $\kappa\xi_b$ is in the
range $\kappa\xi_b \approx 0.85-1$ for
$\phi_0=\phi_c$ \cite{nellen2011}. In preliminary experimental data
\cite{bechinger2011} we find $\kappa\xi_b \approx0.65-0.8$ for
$\phi_0<\phi_c$ and $\kappa\xi_b \approx 1.1-1.8$ for $\phi_0>\phi_c$. These
values are in
the range of those in \figref{fig2} and \figref{fig3}. The difference in
temperature range
between the experiments and our theory can be explained by the different scaling
of
$\xi_b=\xi_0\left|\tau\right|^{-\nu}$. The experimental value of $\xi_0$ and the
critical exponent $\nu=0.61$ \cite{nellen2011} are
quite different from the mean field values of $\xi_0$ and $\nu=0.5$. Thus, the
adsorbed liquid layers in experiments are thicker, leading to attraction at
temperatures higher above $T_c$. Nevertheless, our results qualitatively show
that including ion solvation in the theory accounts for colloid-surface
attraction far from $T_c$ in salt containing mixtures.

\subsection{Interaction between hydrophilic and hydrophobic colloids}

\begin{figure}[!t] 
\includegraphics[width=8cm,clip]{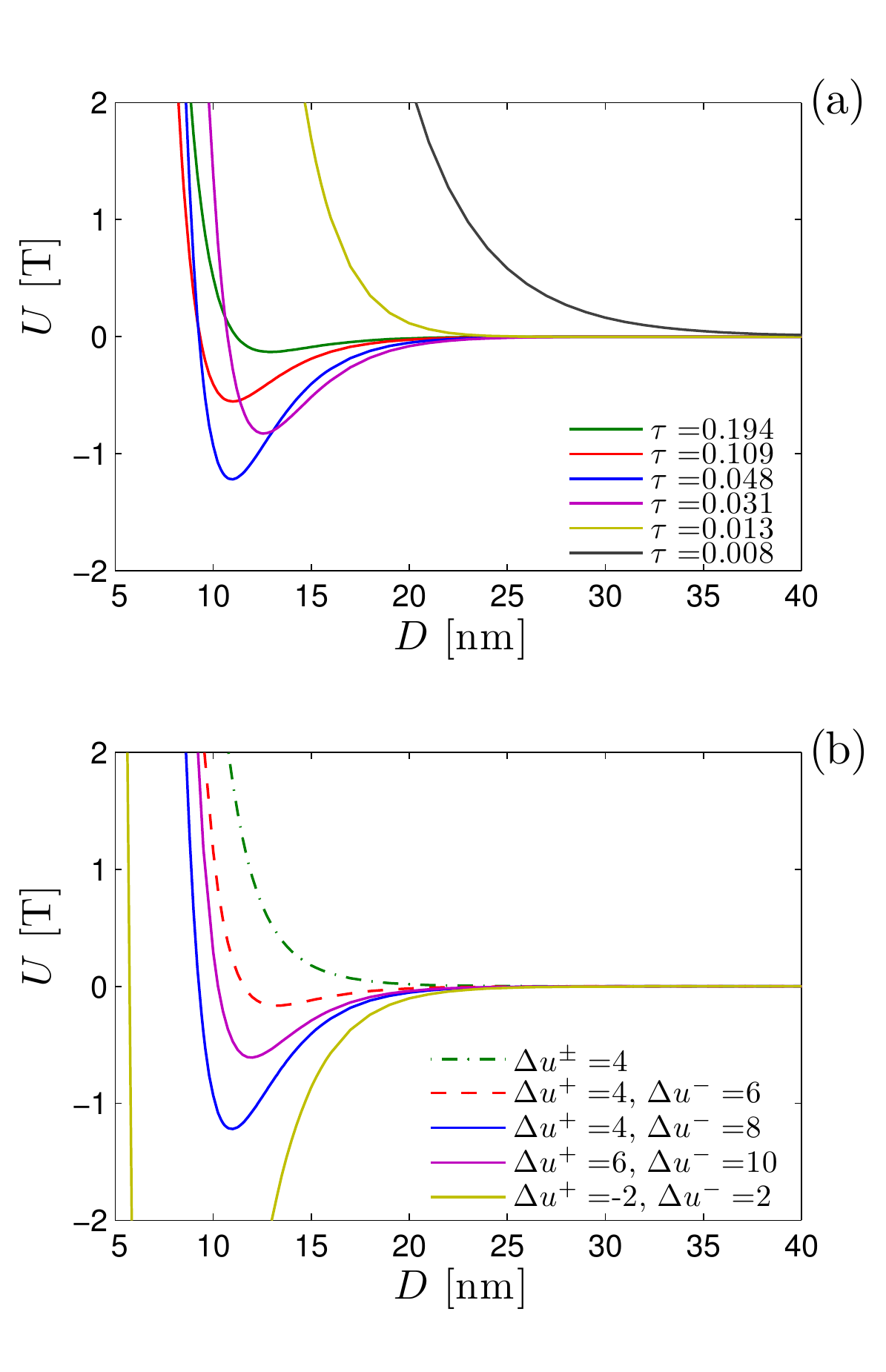}
\caption{ Inter colloid potentials U(D) for hydrophilic and hydrophobic colloids
(antisymmetric boundary conditions). For the surface on the right we used 
$\Delta\gamma_{R}=0.1T/a^2$ and
$\sigma_{R}=-\sigma_{sat}$. For the surface on the left we used
$\Delta\gamma_{L}=-0.4T/a^2$ and $\sigma_{L}=-0.01\sigma_{sat}\ll\sigma_{sat}$. (a)
The interaction potential $U(D)$ at different temperatures $\tau$ showing that $U$ becomes
attractive when $\tau$ decreases, but repulsive close to $T_c$. Here we took for the
ions $\Delta u^+=4$ and $\Delta u^-=8$.
(b) $U(D)$ at $\tau=0.048$ and different values of $\Delta u^\pm$. The
interaction is purely repulsive for $\Delta u^d=\Delta u^+-\Delta u^-=0$
(dash-dot curve) and weakly attractive for $\Delta u^d=2$ (dashed curve). The
attraction is much stronger in the solid curves, all having  different values of
$\Delta u^\pm$ but the same difference $\Delta u^d=-4$. Among these curves, the
attraction is strongest for the antagonistic salt ($\Delta u^-=-\Delta u^+=2$).}
\label{fig5}
\end{figure}

Another physically relevant scenario is that of hydrophilic and hydrophobic
colloids, \textit{i.e} antisymmetric short-range chemical boundary conditions.
For this case the adsorption force in the absence of salt is repulsive
\cite{Fisher1981,Krech1997,Gambassi2009}. However, experimentally it was
observed that in the presence of salt the interaction potential becomes
attractive when the temperature is decreased toward $T_c$, but then repulsive
again as $T_c$ is further approached \cite{nellen2011}. In order to explain this
phenomenon, Pousaneh and Ciach \cite{ciach2011} assumed ions that are insoluble
in the cosolvent and suggested that the attraction is due to the hydration of
ions and is of entropic origin. Another explanation was given by Bier
\textit{et. al.} \cite{bier2011}, who attributed the attraction to the
difference in preferential solvation of ions $\Delta u^d$. In their theory the
attraction is linear in $\Delta u^d$.

\figref{fig5} (a) shows the non-monotonous behavior of $U(D)$ for antisymmetric
surface fields within our theoretical framework. Here we assumed the ions have
different solubilities, $\Delta u^d=-4$, and the surface charge of the
hydrophobic surface is much smaller than that of the hydrophilic surface, as is
usually the case. For these parameters linear theory, relying on the
asymmetry in the solvation energy of ions, 
reproduces surprisingly well the non-monotonous behavior 
\cite{bier2011}. In the nonlinear
regime $\Delta u^{d}\neq0$ is a requisite for this trend but the interaction
depends on the nominal values of $\Delta u^\pm$ as is shown below. 

The onset of the repulsive force is at a surface distance
$D\approx 4\xi_b$, occurring when the two adsorbed solvent layers begin to overlap
\cite{Kiselev2004}. At larger surface separations, the depletion of the more
hydrophilic anions close to the hydrophobic surface gives rise also to an
enhanced and energetically favorable hydrophilic solvent depletion. The net
result is an attractive interaction at $D\gtrsim 4\xi_b$. As the critical
temperature is approached and $\xi_b$ increases the repulsive adsorption force
dominates this relatively small attraction.

\figref{fig5} (b) demonstrates the influence of the value of $\Delta u^+$,
$\Delta u^-$, and their difference $\Delta u^d$ on the interaction profile
$U(D)$. In the solid blue curve we plot $U(D)$ for
$\Delta u^d=-4$ where $\Delta u^+=4$ as in \figref{fig5} (a). Reduction of
$\Delta u^d$ from $-4$ to $-2$ (dashed curve) diminishes the strength of
interaction by an order of magnitude, a strong nonlinear response. For $\Delta
u^d=0$ (dash-dot curve) the interaction is repulsive. \figref{fig5} (b)
shows that the nominal values of $\Delta u^+$ and $\Delta u^-$ are important,
not just their difference. The solid curves in this figure all have $\Delta
u^d=-4$ but different values of $\Delta u^\pm$. For a more hydrophilic salt
(purple curve, $\Delta u^+=6$) the attraction is weaker while for an
antagonistic salt where the cations are hydrophobic and the anions are
hydrophilic (yellow curve, $\Delta u^+=-2$) the attraction is much stronger. The
reason for this is that hydrophobic cations reduce the water adsorption on the
highly charged hydrophilic surface, thus reducing the repulsive adsorption force
and amplifying the asymmetric solvation effect. The antisymmetric case is an
example of the delicate and complex interplay between solvent adsorption and
electrostatic interactions in confined salty mixtures.

\subsection{Ion specific effects}

\begin{figure}[!t] 
\includegraphics[width=8cm,clip]{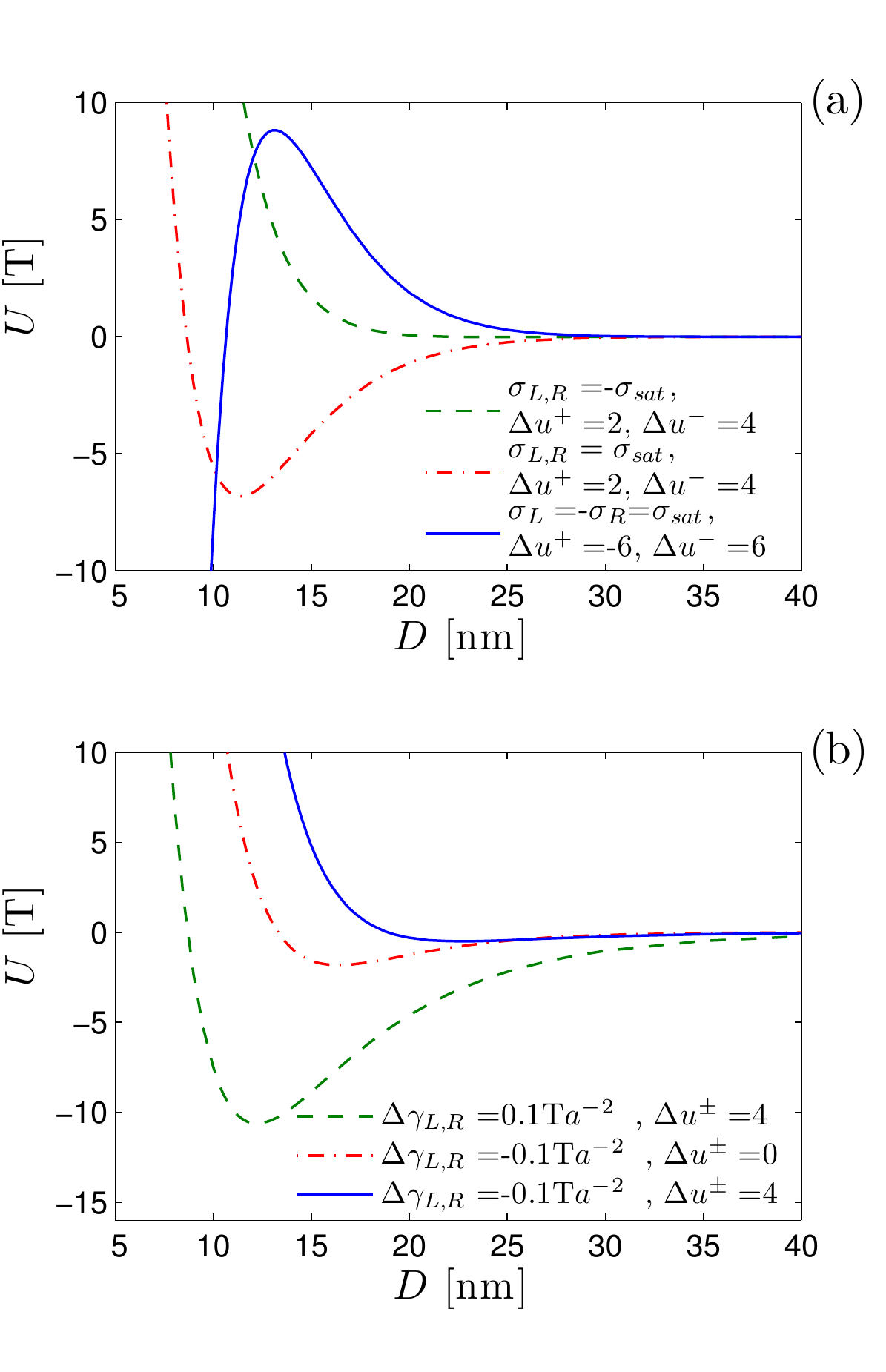}
\caption{(a) The effect of the sign of the colloids' charge 
on the inter-colloid potential $U(D)$. The interaction is attractive for two positively
charged surfaces and is repulsive for two
negatively charged surfaces; compare the dash-dot and dashed curves. We used $\Delta
u^+=2$, $\Delta u^-=4$, $\Delta\gamma_{R,L}=0.1T/a^2$ and $\tau=0.008$. In the solid curve
the surfaces
are both hydrophilic ($\Delta\gamma_{R,L}=0.1T/a^2$) but oppositely charged. For an
antagonistic salt with $\Delta u^-=-\Delta u^+=6$ there is a repulsive regime at
an intermediate range. (b)
The effect of the hydrophilicty or hydrophobicity of the colloid's surface.
In the
absence of preferential solvation ($\Delta u^\pm=0$) two hydrophobic (and
hydrophilic, not shown) surfaces weakly attract (dash-dot curve). For a
hydrophilic salt ($\Delta u^\pm=4$), hydrophobic surfaces repel (solid curve)
whereas hydrophilic surfaces attract (dashed curve). We used $\tau=0.003$
and $\sigma_{L,R}=-\sigma_{sat}$.}
\label{fig6}
\end{figure}

In this section we discuss some consequences of the specific nature of the
ion solvation. The ion solvation energy can vary greatly depending on the
chemical nature of the ion and solvent, its value being typically in the range
$\Delta u \sim 1-10T$. Thus, the influence of the ion-solvent coupling on the
interaction potential is highly ion-specific. In \figref{fig6} (a) we plot
$U(D)$ for hydrophilic ions and surfaces, $\Delta u^+=2$, $\Delta u^-=4$ and
$\Delta\gamma_{R,L}=0.1T/a^2$. When the surfaces are both positively charged
(dash-dot curve), the anions are in excess between the plates and $U(D)$ is
attractive since $\Delta u^-=4$ is large enough and the solvation-related force
overcomes electrostatic repulsion. However, this is not the case for negatively
charged surfaces (dashed curve) where $U(D)$ is repulsive. In this case the
cations are in excess between the plates and $\Delta u^+=2$ is not large enough
to overcome the repulsion. Thus, the difference in the preferential solvation of
cations and anions can produce a selective interaction with respect to the sign of the
surface charge.

An even more intriguing scenario is that of oppositely charged surfaces, where
for hydrophilic surfaces the interaction is expected to be purely attractive
since both adsorption and electrostatic forces are attractive. Nonetheless, if
the mixture contains a strongly antagonistic salt ($\Delta u^+ \Delta
u^-<0$, $\left|\Delta u^d\right| \gg 1$), $U(D)$ has a repulsive region, as is
seen in the solid curve in \figref{fig6} (a) for which $\Delta u^-=-\Delta
u^+=6$. In this curve, since $\sigma_{R}=-\sigma_{L}$ the adsorption of both
anions and cations is significant and similar in magnitude. Hence, since $\Delta
u^-=-\Delta u^+$ the contribution of the solvation energy to the interaction is
small. In addition, at separations $D\gtrsim 4\xi_b$, prior to the merging of
the adsorbed solvent layers \cite{Kiselev2004}, there is significant solvent
depletion near a positively charged wall due to hydrophobic anions. This
depletion reduces the adsorption force and the overall result is a repulsive
interaction. At small separations the adsorption layers merge, $\Gamma$
increases, and combined with the electrostatic attraction the potential is
strongly attractive as expected.

Similar to the effect of the sign of the surface charge, the ion-solvent
coupling renders the interaction specific also relative to the sign of the
surface field. In the absence of salt, if both surfaces are hydrophobic or
hydrophilic they attract \cite{Fisher1978,Fisher1981,Krech1997,Gambassi2009}. As
we saw before, for hydrophilic surfaces the preferential solvation of
hydrophilic ions greatly enhances the interaction far from $T_c$, see the dashed
curve in \figref{fig6} (b) for $\Delta u^\pm=4$ and
$\Delta\gamma_{R,L}=0.1T/a^2$. By reversing the sign of $\Delta\gamma_{R,L}$ 
[solid curve in \figref{fig6} (b)] the interaction becomes repulsive
for hydrophobic surfaces and hydrophilic ions. The dash-dot curve in
\figref{fig6} (b) shows that for hydrophobic surfaces in the absence of
preferential solvation ($\Delta u^\pm=0$) the interaction is weakly attractive.
Here, the addition of hydrophilic ions renders the interaction repulsive by
reducing the adsorption of the solvent on the hydrophobic surface while greatly
increasing the adsorption and attraction for hydrophilic surfaces.

\section{Conclusions}
\label{sec_conclusions}

In summary, we calculated the interaction potential $U(D)$ between two 
charged colloids in salt-containing binary mixtures. The preferential adsorption
of one of the solvents on the colloid's surface combined with the preferential solvation
of the ions gives rise to a strong attraction between the colloids
far from the coexistence curve of the mixture. For surfaces with symmetric chemical
affinity, the interaction is governed by the individual ion solvation ($\propto
\Delta u^\pm$) and to a lesser extent by the difference in solvation energies of
cations and anions.
This is true also for the attraction for antisymmetric boundary
conditions albeit here $\Delta u^d\neq0$ is a requisite for an attractive
potential. 

The ion and solvent densities near the surface of the colloid are highly sensitive to the
ion
solvation energy. Thus, the bulk mixture composition $\phi_0$ and salt
concentration $n_0$ play an important role in determining the inter-colloid 
potential (\figref{fig3} and \figref{fig4}) and this suggests the
possibility of fine-tuning the potential with these readily
controllable parameters. In addition, the ion specific nature of the solvation
energy renders the interaction potential sensitive to the sign of the surface
charge or surface fields, as is shown in \figref{fig6}. We show that 
the interplay between surface and ion-specific interactions may lead to non trivial
effects whereby (i)  two repulsive interactions combine to an attraction
[\figref{fig2} (a)], (ii) for a hydrophilic salt, hydrophobic surfaces repel whereas
hydrophilic surfaces attract [\figref{fig6} (b)], (iii) two oppositely
charged colloids repel down to distances $\sim 10$nm and have a high repulsive barrier of
$5$-$10$T [\figref{fig6} (a)], or (iv) two similarly charged colloids repel when they
are negatively charged and attract when they are both positive [\figref{fig6} (a)].

Our mean-field theory allows the semi-quantitative interpretation of recent
experiments performed in salty mixtures not too close to the critical
temperature \cite{nellen2011,bechinger2011}. We do not look at the influence of
the critical fluctuations on the interaction between surfaces
\cite{Bechinger2008,Fisher1978,Fisher1981,Krech1997,Gambassi2009} although it
has been argued that the addition of salt does not alter the universal critical
behavior of the solvent \cite{bier2011}. 

Recently, multilamellar structures were observed in a bulk binary mixture upon
the addition of an antagonistic salt \cite{Sadakane2009}. In light of the
current work, we believe that the effect of antagonistic salts on the
interaction between \textit{surfaces} is intriguing and deserves similar attention.
In addition, for temperatures below the critical temperature and/or near the
wetting and prewetting transitions, first order capillary condensation
\cite{efips_epl} and bridging transitions \cite{onuki2011} have been predicted.
In light of the current work and these recent findings we stress the importance of
preferential solvation in salty mixtures. Since the Gibbs transfer energy of ions is
usually larger than the thermal energy, the solvation-related force can either induce
large composition perturbations by itself \cite{efips_epl} or amplify them
significantly if already present, as in this work.

\begin{acknowledgments}
We gratefully acknowledge numerous discussions with D. Andelman, C. Bechinger,
M. Bier, J. Dietrich, L. Helden, O. Nellen, A. Onuki and H. Orland. This work was
supported by the Israel Science Foundation under grant No. 11/10 and the
European Research Council ``Starting Grant'' No. 259205.
\end{acknowledgments}

\end{document}